\documentclass[twocolumn,superscriptaddress,amsmath,amssymb]{revtex4-2}

\usepackage{braket,mathtools, hyperref}
\usepackage{upgreek}

\renewcommand{\tilde}{~}
\newcommand{\id}{1\!\!1}
\newcommand{\kluru}{\ket{L\uparrow,R\uparrow}}
\newcommand{\klurd}{\ket{L\uparrow,R\downarrow}}
\newcommand{\kldru}{\ket{L\downarrow,R\uparrow}}
\newcommand{\kldrd}{\ket{L\downarrow,R\downarrow}}

\newcommand{\kaubd}{\ket{A\uparrow,B\downarrow}}
\newcommand{\kadbu}{\ket{A\downarrow,B\uparrow}}

\newcommand{\rlr}{\rho_{\textrm{LR}}}

\newcommand{\rd}{\rho_\textrm{D}}

\newcommand{\kom}{\ket{1_-}}

\newcommand{\usingle}{U^{(1)}}

\begin{document}

    \title{Generating indistinguishability within identical particle systems: spatial deformations as quantum resource activators}

    \author{
        Matteo Piccolini$^{1,2}$,
        Farzam Nosrati$^{1,2}$,
        Gerardo Adesso$^{3}$,
        Roberto Morandotti$^{2}$
        and
        Rosario Lo Franco$^{1}$
        }

        \address{$^{1}$Dipartimento di Ingegneria, Universit\`{a} di Palermo, Viale delle Scienze, 90128 Palermo, Italy\\
        $^{2}$INRS-EMT, 1650 Boulevard Lionel-Boulet, Varennes, Qu\'{e}bec J3X 1S2, Canada\\
        $^{3}$School of Mathematical Sciences and Centre for the Mathematics and Theoretical Physics of Quantum Non-Equilibrium Systems, University of Nottingham, University Park, Nottingham NG7 2RD, United Kingdom}




    \begin{abstract}
        Identical quantum subsystems can possess a property which does not have any classical counterpart: indistinguishability. As a long-debated phenomenon, identical particles' indistinguishability has been shown to be at the heart of various fundamental physical results.
When concerned with the spatial degree of freedom, identical constituents can be made indistinguishable by overlapping their spatial wave functions via appropriately defined \textit{spatial deformations}.
        By the laws of quantum mechanics, any measurement designed to resolve a quantity which depends on the spatial degree of freedom only and performed on the regions of overlap is not able to assign the measured outcome to one specific particle within the system. The result is an entangled state where the measured property is \textit{shared} between the identical constituents.
In this work, we present a coherent formalization of the concept of deformation in a general $N$-particle scenario, together with a suitable measure of the degree of indistinguishability. We highlight the basic differences with nonidentical particles scenarios and discuss the inherent role of spatial deformations as entanglement activators within the \textit{spatially localized operations and classical communication} operational framework.
    \end{abstract}

    \maketitle

    \section{Introduction: identity and indistinguishability}
    \label{section1}

    In physics, particles are said to be \textit{identical} if their intrinsic physical properties, such as mass, electric charge, and (total) spin, are the same\tilde\cite{cohen2006quantum,ghirardi2002entanglement}. This is the case, for example, of subatomic particles such as electrons, photons, quarks, of atomic nuclei, and of atoms and molecules themselves. Particles identity is a cornerstone of both classical and quantum physics which provides the core of the inductive approach to the investigation of Nature's fundamental laws: the assumption that all the electrons in the universe possess the same electric charge, mass, spin, etc., allows to conclude that some fundamental properties extrapolated from the behaviour of a sample of electrons observed in a laboratory also hold for all the other electrons in the universe.

    Despite being frequently used as synonyms, particles identity is not the same as particles \textit{indistinguishability}. Being a purely quantum phenomenon, the latter is more strictly related to the concept of \textit{individual addressability}\tilde\cite{tichyFort,nolabelappr}.
    Identical particles can indeed still be distinguished one from the other when their extrinsic properties, such as their position or the projection of their angular momentum along an axis, are different. This is clear in the classical world where two physical systems, even when microscopic and identical, always occupy distinct positions in space at a fixed time, thus always being potentially individually addressed by following their trajectory\tilde\cite{ghirardi2002entanglement}. On the contrary, this is not always true in quantum mechanics, where the wave-like and probabilistic description of physical systems allows different particles wave functions to be spatially overlapped, thus having a nonzero probability of simultaneously occupying the same region of space. When this situation occurs, any measurement of quantities depending only on the particles position performed on the region of overlap does not allow the observer to understand to which specific particle the measured outcome belongs to. This is the case, for example, of two synchronized photon sources $A$ and $B$ emitting single photons impinging, with a certain probability, on a restricted detecting spatial region. If a single photon detector in that region clicks, we now have no way of knowing from which source the detected photon is coming from: in this situation, we say that there is \textit{no which-way information} and the interested particles are said to be \textit{indistinguishable}\tilde\cite{cohen2006quantum,nolabelappr}.

    The difference between identity and indistinguishability is particularly evident in the everyday experience. It is indeed this difference which allows one to relate observed results to specific samples in an experiment: for example, we can talk about the characterization of a specific laser source carried out in a laboratory in Buenos Aires only because the photons emitted by such a source are very well distinguishable (not spatially overlapped, in this case) from the ones emitted by a neon sign in Tokyo, despite all the photons being identical\tilde\cite{Herbut_1987,tichy,tichyFort,Peresbook}. Still, the laser must be very well isolated from other light sources to be sure that the characterized device is the laser and not a street lamp nearby. Thus, differently from particle identity, particle indistinguishability depends on the variable degrees of freedom involved. As a crucial consequence, indistinguishability is a meaningful concept only when related to the discrimination capability of the measurement device employed to probe those degrees of freedom.

    To better clarify this point, let us recover the above mentioned example of two synchronized single photon emitters and let us now assume that source $A$ is known to emit photons with horizontal polarization, while source $B$ produces only vertically polarized ones. Furthermore, let us suppose that the polarization is not changed by the dynamics. If the single photon detector placed on the region of spatial overlap is designed to discriminate also the photon polarization, we now have a way to understand whether the origin of the particle causing the click is source $A$ or $B$. In other words, the two photons can now be individually addressed and are not indistinguishable anymore despite being identical and spatially overlapped.
    Similarly, if we now further assume the polarization of the two photons to be the same, we could employ a measurement device capable of detecting their energy to discriminate among them. Even the emission time can be used to discriminate between the two particles if we know one source to emit before the other. Finally, the number of detectors can be set to distinguish the two particles, too. For example, let us consider that a photon emitted from source $A$ can only reach regions L and C while a photon emitted from source $B$ can impinge only on C and R, with L, C, and R distinct: a single-photon detector placed in region C would be unable to distinguish the two particles, while the addition of a second detector on L would be enough to reconstruct the origin of every click.

    Summing up, particles are always assumed, often implicitly, to be (or not to be) indistinguishable \textit{to the eyes of the employed measurement devices}, while they are universally identical or nonidentical.
    From the experimental point of view, the actual generation of indistinguishable photons is actually a hard operation of fine tuning and synchronization.
    From now on, we will always implicitly refer to \textit{spatial indistinguishability} when not otherwise specified, i.e. to the indistinguishability of particles spatially overlapped in relation to detectors for which no which-way information exists.

    In this paper we characterize the degree of indistinguishability in a general $N$-particle quantum system. This is achieved by formalizing and extending the idea of \textit{deformation} operations. Firstly introduced in Ref.\tilde\cite{indistdynamicalprotection} and later exploited in Refs.\tilde\cite{Piccolini_2021,piccolini2022indistinguishability} in the particular scenario of bipartite systems, deformations provide a mathematical framework suitable to describe the manipulation of identical constituents when particles' indistinguishability is involved. They account for processes where indistinguishability is generated starting from identical, yet distinguishable particles, and vice versa. Remarkably, they play a fundamental role in devising a coherent extension of the traditional \textit{local operations and classical communication} (LOCC) framework to systems of indistinguishable constituents, whereas the latter fails due to resorting on particles' individuality.
    After a short summary of the no-label approach to identical particles\tilde\cite{nolabelappr,compagno2018dealing} in Section\tilde\ref{section2}, we introduce, formalize, and generalize deformations in Section\tilde\ref{sec:deformation}. In Section\tilde\ref{sec:entropicmeasure}, we retrieve the definition of an entropic measure of spatial indistinguishability firstly introduced in Ref.\tilde\cite{indistentanglprotection}, extending it to the multipartite scenario and to a general amount of degrees of freedom. Finally, in Section\tilde\ref{slocc}, we review and employ the \textit{spatially localized operation and classical communication} (sLOCC) operational framework, which highlights the importance of spatial deformations as a fundamental tool for the manipulation of identical constituents in many practical applications, as confirmed by recent experiments.

    \section{The no-label formalism}
    \label{section2}

    As is well known, particles living in a 3-dimensional space can be divided into two macro groups: bosons, with integer spin, and fermions, with semi-integer spin. According to the symmetrization postulate, the global state describing an ensemble of identical bosons must remain the same when the role of any pair of particles is exchanged: bosonic states are symmetric under particles swapping. On the contrary, fermionic states are ruled to be anti-symmetric under analogous particles exchange\tilde\cite{Peresbook}. The existence of such a postulate is at the heart of the Pauli exclusion principle and sets the ground for fundamental results in modern physics, from models to analyze Bose-Einstein condensates to the description of the behaviour of neutron stars.

 To deal with these conditions, the standard approach to identical particles assigns unphysical (unobservable) labels to each constituent, ensuring that the global state exhibits the correct symmetry when any two labels are switched\tilde\cite{cohen2006quantum}.
For example, let us consider two non-entangled particles with spatial wave functions $\psi_1$, $\psi_2$. If the two particles are nonidentical, their global state is simply given by the tensor product $\ket{\Uppsi^{(2)}}=\ket{\psi_1}_A\otimes\ket{\psi_2}_B$, where the labels $A$ and $B$ encompass all the other physical degrees of freedom as well as the properties which makes the two constituents different. Differently, if the two particles are identical and indistinguishable, labels $A$ and $B$ becomes simply fictitious names without any physical meaning and the global state must be written as\tilde\cite{Peresbook}
    \begin{equation}
    \label{standardstate}
        \ket{\Uppsi^{(2)}}=\frac{1}{\sqrt{2}}\left(\ket{\psi_1}_A\otimes\ket{\psi_2}_B+\eta\,\ket{\psi_2}_A\otimes\ket{\psi_1}_B\right),
    \end{equation}
    in order to satisfy the symmetrization postulate, where $\eta=1$ for bosons and $\eta=-1$ for fermions.

    The approach leading to Eq.\tilde\eqref{standardstate}, despite being the most frequently used even in didactic textbooks, is know to be affected by some formal problems\tilde\cite{tichy,ghirardi}. For example, the necessity to symmetrize/antisymmetrize states by hand as in Eq.\tilde\eqref{standardstate} leads to the emergence of fictitious entanglement when this is evaluated using standard tools such as the von Neumann entropy of the reduced density matrix. This is tackled by adopting ad hoc treatments to probe the existence of quantum correlations among identical particles systems. In addition, such methods require to treat bosons and fermions differently. In order to overcome these problems, a plethora of alternative approaches to deal with identical particles has been proposed over time\tilde\cite{Li2001PRA,Paskauskas2001PRA,cirac2001PRA,zanardiPRA,ghirardi,morrisPRX,eckert2002AnnPhys,balachandranPRL,facchiIJQI,sasaki2011PRA,bose2002indisting,bose2013,tichyFort,killoran2014extracting,sciaraSchmidt,nolabelappr,compagno2018dealing,slocc}.

\begin{figure*}[t!]
		    \centering
		    \includegraphics[width=0.99\textwidth]{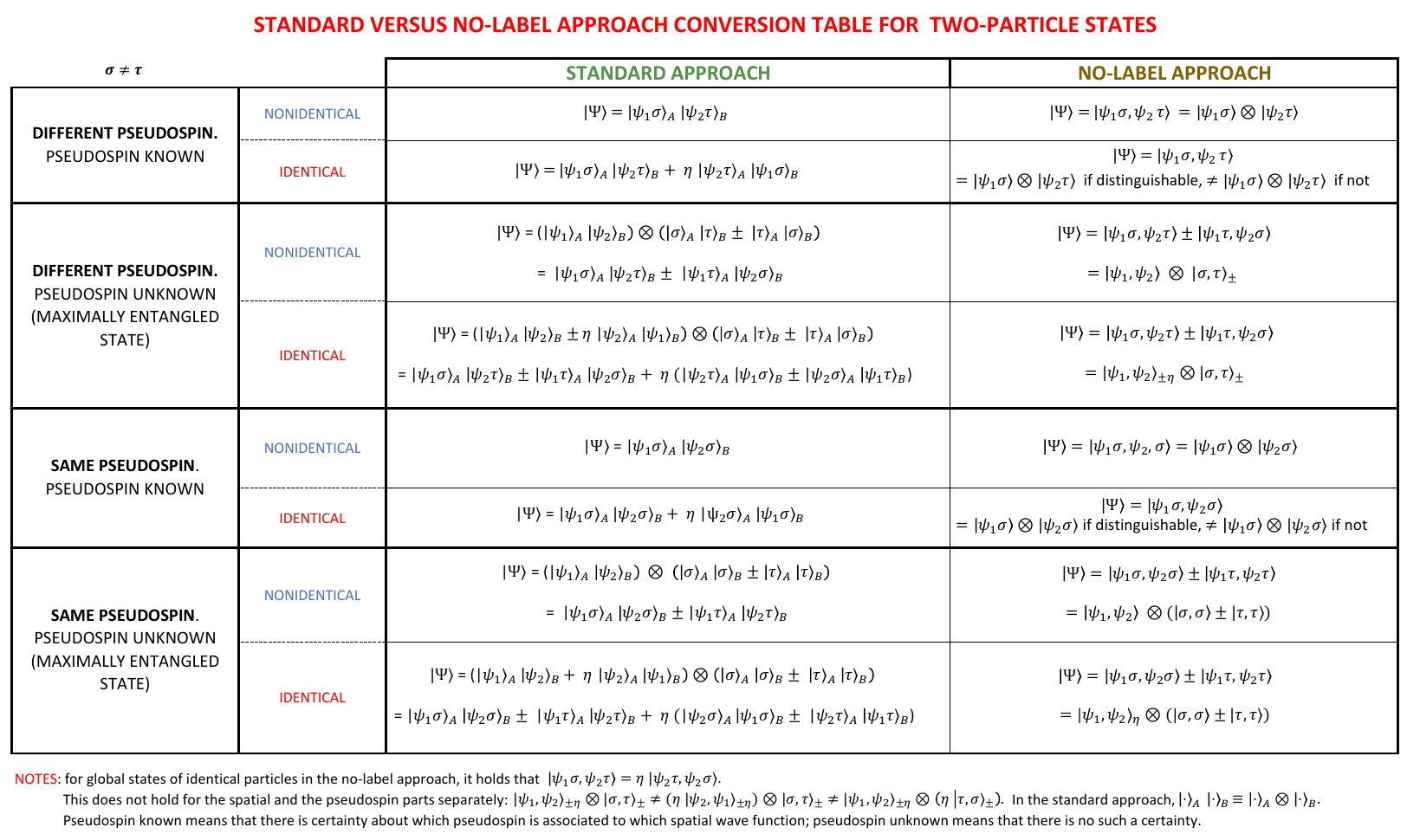}
		    \caption{Conversion table for two-particle states between the standard formalism and the no-label approach. $\psi_1$ and $\psi_2$ are the two single particle spatial wave functions, while $\sigma$ and $\tau$ ($\sigma\neq\tau$) are the pseudospin projection along a preferred axis. Notation is reported for both nonidentical and identical particles states: for the first ones, labels used in the standard approach have a physical meaning, identifying physical, measurable properties; for the latter, no physical meaning can be assigned to labels when the described particles are indistinguishable. The no-label approach overcomes this problem by avoiding to resort on labels. The reported structures of states of identical particles is preserved when exchanging the roles of the spatial (external) and pseudospin (internal) degrees of freedom. Normalization coefficients are omitted to avoid cluttering.}
		\label{conversion}
	    \end{figure*}

    Among these methods, the \textit{no-label approach} recognizes the origin of the problem in the unphysical labels $A$ and $B$ appearing in Eq.\tilde\eqref{standardstate}, removing them from the formalism\tilde\cite{nolabelappr,compagno2018dealing}. In this way, global states are simply given by a list of the single particle states: considering once again the example of two constituents with single spatial wave functions $\psi_1$ and $\psi_2$, the global state is written as $\ket{\Uppsi^{(2)}}:=\ket{\psi_1,\psi_2}$. If the two particles are distinguishable, e.g. not spatially overlapped, the global state is still a product state. Nonetheless, when they are not perfectly distinguishable, the global state cannot be written as a tensor product anymore: $\ket{\Uppsi^{(2)}}\neq\ket{\psi_1}\otimes\ket{\psi_2}$. Similarly, the global Hilbert space $\mathcal{H}^{(2)}$ is generally not the tensor product of the single particle Hilbert spaces $\mathcal{H}^{(1)}_1$ and $\mathcal{H}^{(1)}_2$: $\mathcal{H}^{(2)}\neq\mathcal{H}^{(1)}_1\otimes\mathcal{H}^{(1)}_2$. The generalization to the $N$-particle scenario is straightforward, with the global state $\ket{\Uppsi^{(N)}}:=\ket{\psi_1,\psi_2,\dots,\psi_N}$ generally satisfying $\ket{\Uppsi^{(N)}}\neq\ket{\psi_1}\otimes\ket{\psi_2}\otimes\dots\otimes\ket{\psi_N}$. For a more exhaustive review of how the most common two-particle states are written within the no-label approach, compared to their expressions in the standard approach with fictitious labels, see the table in Fig.\tilde\ref{conversion}.

    Given the two-particle state $\ket{\psi_1,\psi_2}$, the cornerstone of the no-label approach is provided by the definition of the probability amplitude related to finding the system in the state $\ket{\varphi_1,\varphi_2}$, which takes into account the eventual indistinguishability of the constituents. According to the meaning of indistinguishability discussed in Section\tilde\ref{section1}, the impossibility to discriminate between the two particles should reasonably lead to both of them contributing with their probability amplitude of being found in $\varphi_1$ and $\varphi_2$. Thus, we define
    \begin{equation}
    \label{nolabelprobampl}
        \braket{\varphi_1,\varphi_2|\psi_1,\psi_2}:=
        \braket{\varphi_1|\psi_1}\braket{\varphi_2|\psi_2}
        +\eta\braket{\varphi_1|\psi_2}\braket{\varphi_2|\psi_1}.
    \end{equation}
    Remarkably, this definition directly encodes the statistical exchange phase $\eta$: within the no-label approach, the statistical information about the identical particles nature is encoded in the transition amplitudes, rather than in the symmetrization of the quantum state.
    Some important characteristics of the formalism can be directly derived from\tilde\eqref{nolabelprobampl}: comparing $\braket{\varphi_1,\varphi_2|\psi_1,\psi_2}$ with $\braket{\varphi_1,\varphi_2|\psi_2,\psi_1}$, it follows that
    \begin{equation}
    \label{nolabelswapping}
        \ket{\psi_2,\psi_1}=\eta\ket{\psi_1,\psi_2}
    \end{equation}
    (see the note at the bottom of Fig.\tilde\ref{conversion}).
    Furthermore, state $\ket{\psi_1,\psi_2}$ is not, in general, normalized: indeed, it can be easily checked that (assuming the single particle wave functions $\psi_1,\,\psi_2$ to be properly normalized)
    \begin{equation}
    \label{nolabelinnerproduct}
       \braket{\psi_1,\psi_2|\psi_1,\psi_2}=1+\eta\,\lvert\braket{\psi_1|\psi_2}\rvert^2:=C_+^2,
    \end{equation}
    implying that the correctly normalized two particle state is
    \begin{equation}
    \label{nolabelnormalized}
        \ket{\Uppsi^{(2)}}_N=\ket{\psi_1,\psi_2}/C_+.
    \end{equation}
    Notice that, when the spatial overlap is null (i.e. distinguishable particles), $\braket{\psi_1|\psi_2}=0$ and the normalized two particle state simply reduces to $\ket{\Uppsi^{(2)}}_N=\ket{\psi_1,\psi_2}$.
    Eq.\tilde\eqref{nolabelprobampl}, Eq.\tilde\eqref{nolabelswapping}, and Eq.\tilde\eqref{nolabelnormalized} can be easily extended to the general $N$-particle scenario: given the states $\ket{\psi_1,\psi_2,\dots,\psi_N}$ and $\ket{\varphi_1,\varphi_2,\dots,\varphi_N}$, the related $N$-particle probability amplitude is given by
    \begin{widetext}
        \begin{equation}
        \label{innerindist}
            \braket{\varphi_1,\varphi_2,\dots,\varphi_N|\psi_1,\psi_2,\dots,\psi_N}
            =\sum_{\vec{\alpha}}\eta^{P_{\vec{\alpha}}}
            \braket{\varphi_1|\psi_{\alpha_1}}\braket{\varphi_2|\psi_{\alpha_2}}\dots\braket{\varphi_N|\psi_{\alpha_N}},
        \end{equation}
    \end{widetext}
    where $\vec{\alpha}=(\alpha_1,\alpha_2,\dots,\alpha_N)$ is any arbitrary permutation of $(1,2,\dots,N)$, while $P_{\vec{\alpha}}$ is the parity of the permutation.
    Under particle swapping, the $N$-particle state behaves as
    \begin{equation}
        \ket{\psi_{\alpha_1},\psi_{\alpha_2},\dots,\psi_{\alpha_N}}
        =\eta^{P_{\vec{\alpha}}}\ket{\psi_1,\psi_2,\dots,\psi_N},
    \end{equation}
    while the properly normalized state is simply given by
    \begin{equation}
        \ket{\Uppsi^{(N)}}_N=\ket{\psi_1,\dots,\psi_N}/
        \sqrt{\braket{\psi_1,\dots,\psi_N|\psi_1,\dots,\psi_N}}.
    \end{equation}
    Notice that, if all the single particle wave functions are non-overlapping and individually normalized (distinguishable scenario), from Eq.\tilde\eqref{innerindist} we have
    \begin{equation}
    \label{innerdist}
        \braket{\psi_1,\dots,\psi_N|\psi_1,\dots,\psi_N}
        =\braket{\psi_1|\psi_1}
        \dots
        \braket{\psi_N|\psi_N}=1
    \end{equation}
    and $\ket{\Uppsi^{(N)}}_N= \ket{\psi_1,\psi_2,\dots,\psi_N}$.


    When dealing with distinguishable particles, it is possible to resort to the \textit{local operations and classical communication} framework to manipulate, quantify, and compare entanglement\tilde\cite{horodecki2009quantum}. Here, \textit{local} refers to the concept of \textit{particle locality} and to the possibility of acting on the single constituents individually. Thus, such an approach is not applicable to systems of indistinguishable particles, where no individual constituent can be defined. In such a situation, one can instead rely on operations which are localized \textit{in space}, rather than on single elements, leading to the \textit{spatially localized operations and classical communication} (sLOCC) framework discussed further in Section\tilde\ref{slocc}\tilde\cite{slocc,experimentalslocc}. Within this scenario, the action of a single particle operator $O^{(1)}_X$ localized on the spatial region $X$ on the multipartite state $\ket{\Uppsi^{(N)}}$ is defined, according to the no-label approach, as
    \begin{equation}
    \label{indistsingleopaction}
        O^{(1)}_X\ket{\Uppsi^{(N)}}
        :=\sum_i\,\lvert\braket{X|\psi_i}\rvert\ket{\psi_1,\dots,O^{(1)}_X\,\psi_i,\dots,\psi_N},
    \end{equation}
    where the presence of at least one constituent in the region $X$ is assumed\tilde\cite{indistdynamicalprotection}.
    Remarkably, the operational necessity of focusing on a specific region of space rather than on individual particles is reflected, in Eq.\tilde\eqref{indistsingleopaction}, by the sum being weighted by the probability amplitudes associated to each particle being in the region $X$. Notice that, when the region $X$ is wide enough to enclose the whole spatial distribution of $\ket{\Uppsi^{(N)}}$, Eq.\tilde\eqref{indistsingleopaction} reduces to
    \begin{equation}
        O^{(1)}_X\ket{\Uppsi^{(N)}}
        :=\sum_i\,\ket{\psi_1,\dots,O^{(1)}_X\,\psi_i,\dots,\psi_N},
    \end{equation}
    which is the usual single-particle operation acting on a state of $N$ identical particles.\newline

    \section{Deformations}
    \label{sec:deformation}

    In this Section, we discuss and formalize the concept of \textit{deformation}, a tool of particular importance when applied to systems of identical particles.

    In contrast to global unitary transformations where all the elements of a multipartite state are modified in the same way, deformations consist in transformations acting differently, but still unitarily, on each particle, thus changing the relative relations among the constituents.
    Given an $N$-partite state $\ket{\Uppsi^{(N)}}=\ket{\psi_1,\psi_2,\dots,\psi_N}$ of either distinguishable or indistinguishable particles, the action of the deformation $D^{(N)}_{\vec{a},\vec{X}}$ is defined, within the no-label approach, as
    \begin{widetext}
        \begin{equation}
        \label{deformation}
            \begin{aligned}
                D^{(N)}_{\vec{a},\vec{X}}\ket{\Uppsi^{(N)}}
                :&=\Big(\usingle_{a_1,X_1}\otimes\usingle_{a_2,X_2}\otimes\dots\otimes\usingle_{a_N,X_N}\Big)\ket{\Uppsi^{(N)}}\\
                &=\sum_{\vec{\alpha}}\,\lvert
                \braket{X_1|\psi_{\alpha_1}}\braket{X_2|\psi_{\alpha_2}}\dots\braket{X_N|\psi_{\alpha_N}}
                \rvert
                \,\,
                \eta^{P_{\vec{\alpha}}}\,
                \ket{\usingle_{a_1,X_1}\psi_{\alpha_1},\usingle_{a_2,X_2}\psi_{\alpha_2},\dots,\usingle_{a_N,X_N}\psi_N}.
            \end{aligned}
        \end{equation}
    \end{widetext}
   Here, the elements $a_j$ in $\vec{a}=(a_1,a_2,\dots,a_N)$ identify the type of transformation represented by the single particle unitary operator $U^{(1)}_{a_j,X_j}$ and encode the set of parameters required to determine it, while $X_j\in\vec{X}=(X_1,X_2,\dots,X_N)$ denotes its region of action. $\vec{\alpha}$ and
    $P_{\vec{\alpha}}$ are as in Eq.\tilde\eqref{innerindist}. In general, for a deformation $a_j\neq a_i$ for $j\neq i$.
    Eq.\tilde\eqref{deformation} holds when each operator acts on at least one particle, i.e. $\exists\, \vec{\alpha}:\,\forall\,i\,\exists\, j:\,\braket{X_i|\psi_{\alpha_j}}\neq0$.

    We define the deformation operator to be linear, that is
    \begin{widetext}
        \begin{equation}
        D^{(N)}_{\vec{a},\vec{X}}\ket{\lambda_1\Uppsi^{(N)}_1+\lambda_2\Uppsi^{(N)}_2}=\lambda_1D^{(N)}_{\vec{a},\vec{X}}\ket{\Uppsi^{(N)}_1}+\lambda_2D^{(N)}_{\vec{a},\vec{X}}\ket{\Uppsi^{(N)}_2},\ \forall \lambda_1,\lambda_2\in \mathbb{C}.
    \end{equation}
    \end{widetext}
    The probability amplitudes weighting the sum in Eq.\tilde\eqref{deformation} account, as in\tilde\eqref{indistsingleopaction}, for the spatially localized approach required when the constituents are indistinguishable. When they are distinguishable, either being identical or nonidentical, we can individually address each of them within the traditional LOCC framework and drop the subscript $X$, so that Eq.\tilde\eqref{deformation} becomes
    \begin{equation}
    \label{deformationdistinguishable}
        D^{(N)}_{\vec{a}}\ket{\Uppsi^{(N)}}
        =\ket{\usingle_{a_1}\psi_1,\usingle_{a_2}\psi_2,\dots,\usingle_{a_N}\psi_N}.
    \end{equation}
    Moreover, deformations are unitary when dealing with nonidentical particles. Indeed, in this case we are sure that the constituents are left distinguishable by the deformation. Thus, the right hand side of Eq.\tilde\eqref{deformationdistinguishable} reduces in this case to a tensor product, namely
    \begin{equation}
    \label{deformationnonid}
        D^{(N)}_{\vec{a}}\ket{\Uppsi^{(N)}}
        =\ket{\usingle_{a_1}\psi_1}
        \otimes
        \ket{\usingle_{a_2}\psi_2}
        \otimes\dots\otimes
        \ket{\usingle_{a_N}\psi_N}.
    \end{equation}
    Hence, one has
    \begin{widetext}
        \[
        \begin{aligned}
            \braket{D^{(N)}_{\vec{a}}\Uppsi^{(N)}|D^{(N)}_{\vec{a}}\Uppsi^{(N)}}
            &=
            \braket{\usingle_{a_1}\psi_1|\usingle_{a_1}\psi_1}
            \braket{\usingle_{a_2}\psi_2|\usingle_{a_2}\psi_2}
            \dots
            \braket{\usingle_{a_N}\psi_N|\usingle_{a_N}\psi_N}
            \\
            &=\braket{\psi_1|\psi_1}
            \braket{\psi_2|\psi_2}
            \dots
            \braket{\psi_N|\psi_N},
        \end{aligned}
        \]
    \end{widetext}
    which implies
    $\bra{\Uppsi^{(N)}}
       \left[
        D^{(N)}_{\vec{a}}
        \right]^\dagger D^{(N)}_{\vec{a}}\ket{\Uppsi^{(N)}}
        =\braket{\Uppsi^{(N)}|\Uppsi^{(N)}}$, finally leading to
    \begin{equation}
        \big[
        D^{(N)}_{\vec{a}}
        \big]^\dagger D^{(N)}_{\vec{a}}=\id.
    \end{equation}
    Remarkably, this is in general not true anymore for identical constituents, not even when initially distinguishable. From the physical point of view, this is so because the deformation can change the relative spatial overlap of particles, thus leading to the emergence of indistinguishability manifested in the cross-inner products appearing in the right hand side of Eq.\tilde\eqref{innerindist}. In order to explicitly show this, let us consider the scenario of $N=2$ distinguishable but identical particles for simplicity. Before applying the deformation, from Eq.\tilde\eqref{innerdist} we have
    \[
        \braket{\Uppsi^{(2)}|\Uppsi^{(2)}}
        =\braket{\psi_1,\psi_2|\psi_1,\psi_2}
        =\braket{\psi_1|\psi_1}\braket{\psi_2|\psi_2}.
    \]
    After the deformation, instead, from Eq.\tilde\eqref{innerindist} it holds that
    \begin{widetext}
        \[
        \begin{aligned}
            \braket{D^{(2)}_{\vec{a}}\Uppsi^{(2)}|D^{(2)}_{\vec{a}}\Uppsi^{(2)}}
            &=\braket{U_{a_1}^{(1)}\psi_1,U_{a_2}^{(1)}\psi_2|U_{a_1}^{(1)}\psi_1,U_{a_2}^{(1)}\psi_2}\\
            &=\braket{U_{a_1}^{(1)}\psi_1|U_{a_1}^{(1)}\psi_1}\braket{U_{a_2}^{(1)}\psi_2|U_{a_2}^{(1)}\psi_2}
            +\eta\,
            \lvert\braket{U_{a_1}^{(1)}\psi_1|U_{a_2}^{(1)}\psi_2}\rvert^2\\
            &=\braket{\psi_1|\psi_1}\braket{\psi_2|\psi_2}
            +\eta\,\lvert\braket{\psi_1|\big[U_{a_1}^{(1)}\big]^\dagger U_{a_2}^{(1)}|\psi_2}\rvert^2.
        \end{aligned}
        \]
    \end{widetext}
    Since, in general, $[U_i^{(1)}]^\dagger U_j^{(1)}\neq 0$, it follows that
    \begin{equation}
        \begin{aligned}
            \bra{\Uppsi^{(2)}}
            \big[
            D^{(2)}_{\vec{a}}
            \big]^\dagger D^{(2)}_{\vec{a}}\ket{\Uppsi^{(2)}}
            &\neq
            \braket{\Uppsi^{(2)}|\Uppsi^{(2)}}\Rightarrow
            \big[
            D^{(2)}_{\vec{a}}
            \big]^\dagger D^{(2)}_{\vec{a}}\neq\id.
        \end{aligned}
    \end{equation}

\begin{figure*}[t!]
		    \includegraphics[width=0.87\textwidth]{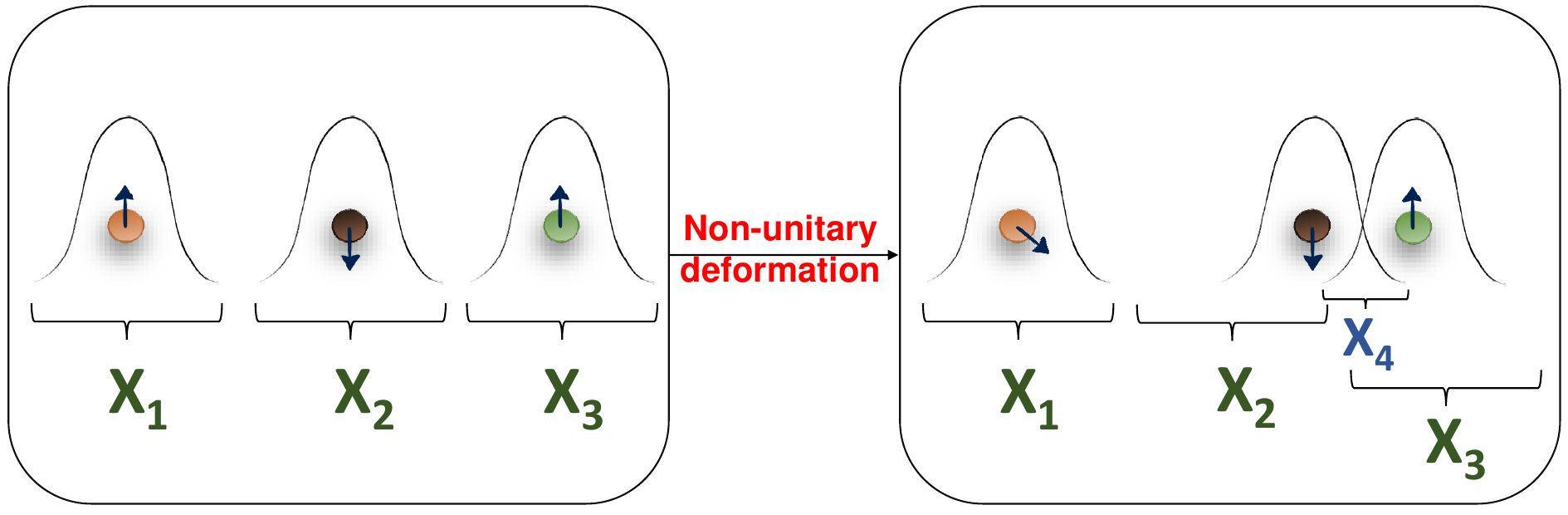} \\
            \includegraphics[width=0.99\textwidth]{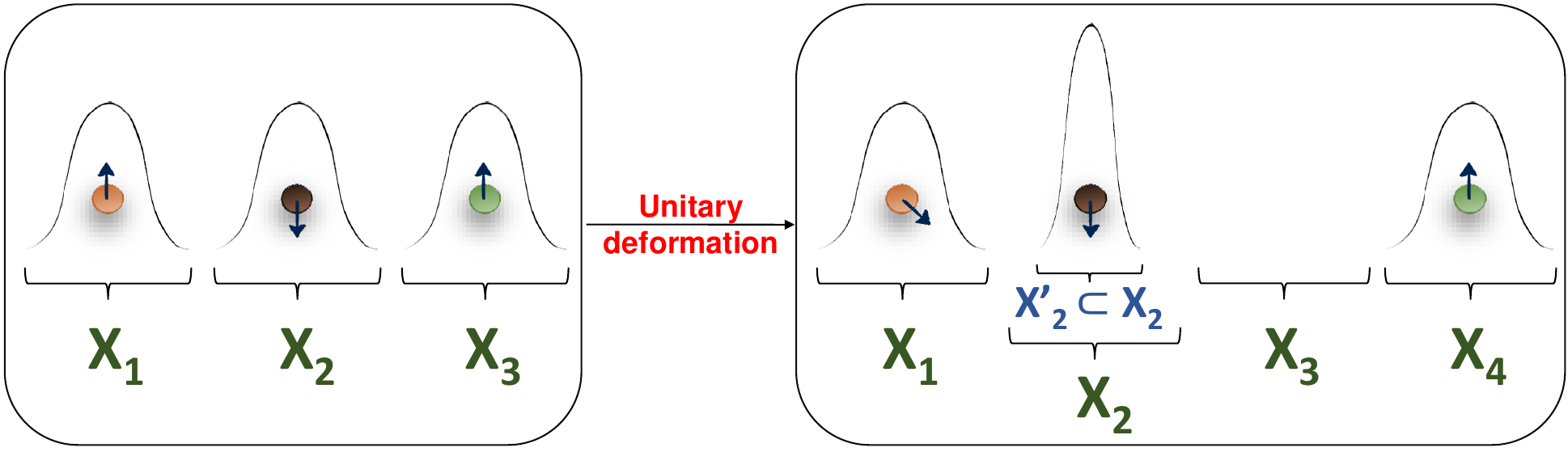}
		    \caption{{\bfseries Top.} Example of non-unitary deformation of three identical and initially distinguishable particles. The particle localized in region $X_1$ undergoes a spin rotation, while the ones in regions $X_2$ and $X_3$ get spatially overlapped over region $X_4$, where spatial indistinguishability is generated. {\bfseries Bottom.} Example of unitary deformation of three identical and distinguishable particles. The particle localized in region $X_1$ undergoes a spin rotation, the one in region $X_2$ sees a unitary restriction of its wave function support to a region $X'_2\subset X_2$, while the particle in region $X_3$ gets spatially translated to $X_4$. No indistinguishability is generated by the process.}
		\label{deformazione2}
	\end{figure*}

    We thus conclude that \textit{deformations are unitary when applied to nonidentical particles and, in general, non-unitary for identical ones}.
    The latter situation is schematically represented in Figure\tilde\ref{deformazione2}(top), where we depict an example of deformation acting on three identical, nonetheless distinguishable, particles leading to the generation of spatial indistinguishability, thus being non-unitary. In Figure\tilde\ref{deformazione2}(bottom), instead, we report a pictorial representation of the particular scenario where three identical, distinguishable particles are manipulated via a deformation which does not generate indistinguishability, thus retaining unitarity.

   Clearly, when normalization is important, states of indistinguishable particles obtained by a deformation can be straightforwardly normalized: given a system of $N$ identical particles in a general mixed state $\rho$, the normalized state after the deformation is
    \begin{equation}
    \label{normaldef}
        \rho_N=\frac{D\rho D^\dagger}{\text{Tr}\left[D^\dagger D\,\rho\right]},
    \end{equation}
    where we omit superscripts and subscripts of the deformation operator for simplicity.

    \section{Entropic measure of indistinguishability}
    \label{sec:entropicmeasure}

    As we shall discuss in the next Section, spatial indistinguishability provides an important quantum resource which can be accessed within the sLOCC operational framework for different goals. Within this picture, deformations generating indistinguishability from previously distinguishable constituents provide the key tool to activate such resources. In order to quantitatively demonstrate this, we need a way to quantify indistinguishability. To this aim, we now introduce en entropic measure of generalized indistinguishability, of which spatial indistinguishability is derived as a particular case.

    Let us consider a general $N$-particle state $\ket{\Uppsi^{(N)}}=\ket{\psi_1,\psi_2,\dots,\psi_N}$; here, each $\psi_j$ encodes both the single particle spatial wave function $\chi_j$ and all the other relevant degrees of freedom given by the eigenvalues of a complete set of commuting observables and gathered in a vector $\vec{\sigma}_j$, so that $\ket{\psi_j}=\ket{\chi_j\,\vec{\sigma}_j}$. We now identify $N$ distinct regions of space $S_1,\,S_2,\dots,S_N$ where we set $N$ single particle detectors, corresponding to the non-overlapping spatial modes $\ket{S_1},\,\ket{S_2},\,\dots,\,\ket{S_N}$. In general, the detectors will be sensible to the spatial position of particles (by construction) and to a subset $\vec{\alpha}$ of the degrees of freedom encoded in $\vec{\sigma}$, while being unable to detect the remaining $\vec{\beta}$ (where $\vec{\sigma}_j=\vec{\alpha}_j\cup\vec{\beta}_j\,\forall\,j=1,\dots,N$). For example, each detector could be capable of detecting the energy of a particle impinging on the spatial region where it is set, without having access to its spin. A single particle detection performed in the region $S_k$ giving as outcomes the set of values $\vec{\alpha}_k$ is thus described by the projection operator $\Pi_{k}^{(1)}=\sum_{\vec{\beta}}\ket{S_k\,\vec{\alpha}_k\,\vec{\beta}}\bra{S_k\,\vec{\alpha}_k\,\vec{\beta}}$, while the probability of such an outcome when detecting a particle whose state is $\ket{\psi_j}=\ket{\chi_j\,\vec{\alpha}_j\,\vec{\beta}_j}$ is given by
    \begin{equation}
        \begin{aligned}
        P_{k,j}
        &=\braket{\psi_j|\Pi_{k}^{(1)}|\psi_j}
        =\sum_{\vec{\beta}}
        \lvert
            \braket{S_k\,\vec{\alpha}_k\,\vec{\beta}|\psi_j}
        \rvert^2\\
        &=\lvert
            \braket{S_k|\chi_j}\braket{\vec{\alpha}_k|\vec{\alpha}_j}
        \rvert^2.
        \end{aligned}
    \end{equation}
    Since the spatial regions $S_k$ are distinct, a global simultaneous detection of the multipartite state giving as outcomes $\vec{\alpha}_1$ for the particle in the region $S_1$, $\vec{\alpha}_2$ for the one in $S_2$, and so on, is described by the action of the $N$-particle projection operator
    \begin{equation}
    \label{measurement}
        \Pi^{(N)}_{\{S_k,\vec{\alpha}_k\}}=\bigotimes_{k=1}^{N}\Pi_k^{(1)}.
    \end{equation}
    We now introduce the joint probability related to the projective measurement in Eq.\tilde\eqref{measurement} of detecting in the region $S_1$ the particle whose state is $\ket{\psi_{j_1}}$, in the region $S_2$ the one whose state is $\ket{\psi_{j_2}}$, and so on, that is
    \begin{equation}
    \label{jointprob}
      P_{\{S_k,\vec{\alpha}_k\}}^{j_1,\dots,j_N}
      =\prod_{k=1}^N P_{k,j_k}.
    \end{equation}
    With respect to the projective measurement in Eq.\tilde\eqref{measurement}, we define the \textit{degree of indistinguishability} of the $N$-particle state as
    \begin{equation}
    \label{indistgeneral}
        \mathcal{I}_{\{S_k,\vec{\alpha}_k\}}
        =-\sum_{\substack{j_1,\dots,j_N=1
        \\j_1\neq\dots\neq j_N}}^N
        \frac{
            P_{\{S_k,\vec{\alpha}_k\}}^{j_1,\dots,j_N}
            }
            {\mathcal{Z}}
        \log_2
        \frac{
            P_{\{S_k,\vec{\alpha}_k\}}^{j_1,\dots,j_N}
            }
            {\mathcal{Z}},
    \end{equation}
    where we have indicated the partition function
    \begin{equation}
    \label{partition}
        \mathcal{Z}
        =\sum_{\substack{j_1,\dots,j_N=1
        \\j_1\neq\dots\neq j_N}}^NP_{\{S_k,\vec{\alpha}_k\}}^{j_1,\dots,j_N}.
    \end{equation}
    When all the particles are spatially separated, there is at most only one non-null joint probability contributing to Eq.\tilde\eqref{indistgeneral}. In particular, if they are perfectly localized on one region each and the values of their accessible degrees of freedom are $\{\vec{\alpha}_k\}_{k=1}^N$, such a probability is equal to 1 and $\mathcal{I}$ reaches its minimum $\mathcal{I}_{\{S_k,\vec{\alpha}_k\}}=0$: particles are perfectly distinguishable with respect to the measurement given by Eq.\tilde\eqref{measurement}. On the contrary, if all the constituents are equally distributed over all the $N$ spatial regions and possess the same values $\vec{\alpha}_1=\vec{\alpha}_2=\dots=\vec{\alpha}_N$, then all the joint probabilities contribute equally to Eq.\tilde\eqref{indistgeneral}: we have maximally indistinguishable particles and $\mathcal{I}$ takes its maximum value $\mathcal{I}_{\{S_k,\vec{\alpha}_k\}}=\log_2 N!$.

    In what follows, we shall be interested in the scenario where the detectors are only sensible to the spatial degree of freedom. This situation is derived from the above described picture by setting $\vec{\alpha}=\{\emptyset\}$, so that Eq.\tilde\eqref{indistgeneral} reduces to a measure of the degree of \textit{spatial} indistinguishability (see Section\tilde\ref{slocc}).

    \section{Accessing quantum indistinguishability resources: the sLOCC operational framework}
    \label{slocc}

    As discussed in Section\tilde\ref{section2}, indistinguishable particles cannot be addressed with the traditional LOCC framework, since this relies on the possibility to individually manipulate, and thus distinguish, the single constituents. From an operational point of view, we thus resort to the sLOCC framework to access the quantum properties of an indistinguishable particles state\tilde\cite{slocc,experimentalslocc}.

    For simplicity, we present the sLOCC framework within the simple scenario of two identical qubits with opposite pseudospin, initially distinguishable and localized in the distinct spatial regions $A$ and $B$. Following the original formulation\tilde\cite{slocc}, we take the bipartite system to be in the initial state $\ket{\Uppsi}_\textrm{AB}=\ket{A\uparrow,B\downarrow}$. Notice that $\ket{\Uppsi}_\textrm{AB}$ is normalized, since $\braket{A|B}=0$.
    Applying the notions introduced in Section\tilde\ref{sec:deformation}, we proceed by deforming such a state to make the two single particle wave functions spatially overlap over two distinct regions $L$ and $R$ corresponding to the normalized spatial modes $\ket{L},\,\ket{R}$. This amount to performing the transformation
    \begin{equation}
    \label{deformedslocc}
        \ket{\Uppsi}_\textrm{AB}=\ket{A\uparrow,B\downarrow}
        \overset{D}{\longrightarrow}
        \ket{\Uppsi}_D=\ket{\psi_1\uparrow,\psi_2\downarrow},
    \end{equation}
    where $\ket{\psi_1}=l\ket{L}+r\ket{R}$ and $\ket{\psi_2}=l'\ket{L}+r'\ket{R}$. Here, the complex coefficients $l,l',r,r'$ determine the different probabilities of finding each particle in each region and satisfy the relation $|l|^2+|r|^2=|l'|^2+|r'|^2=1$.
    Following what discussed in Section\tilde\ref{section1} we highlight that, despite being spatially indistinguishable, the two qubits in state $\ket{\Uppsi}_D$ can still be discriminated by a device capable of detecting their spin direction, which has been left unchanged by the deformation. Finally, the deformation has left the state normalized: indeed, it holds that
    \begin{widetext}
        \[
        \prescript{}{D}{\braket{\Uppsi|\Uppsi}_D}
        =\Big(\braket{\psi_1|\psi_1}\braket{\uparrow|\uparrow}\Big)
        \Big(\braket{\psi_2|\psi_2}\braket{\downarrow|\downarrow}\Big)
        +\eta\,\lvert\braket{\psi_1|\psi_2}\braket{\uparrow|\downarrow}\rvert^2=1.
        \]
    \end{widetext}
    We now set two single-particle detectors on $L$ and $R$ respectively and perform a coincidence measurement, preserving the state if both of them detect a particle and discarding it otherwise. Crucially, the detectors are unable to access the spin direction, so that the two qubits are effectively indistinguishable to their eyes. Thus, this part of the process amounts to a postselected measurement where state $\ket{\Uppsi}_D$ is projected on the subspace spanned by the basis
    \begin{equation}
	\label{basis}
	    \mathcal{B}_\textrm{LR}=\{\kluru,\klurd,\kldru,\kldrd\}
	\end{equation}
    via the corresponding projection operator
    \begin{equation}
	\label{sloccprojector}
    	\Pi_\textrm{LR}
    	=\sum_{\sigma,\tau=\uparrow,\downarrow}\ket{L\sigma,R\tau}\bra{L\sigma,R\tau}.
	\end{equation}
	After the proper normalization, the resulting state is given by
	    \begin{eqnarray}
	    \label{sloccstatepure}
    	    \ket{\Uppsi}_\textrm{LR}
    	    &=&\frac{\Pi_\textrm{LR}\ket{\Uppsi}_D}{\sqrt{\prescript{}{D}{\braket{\Uppsi|\Pi_\textrm{LR}|\Uppsi}_D}}}\nonumber\\
    	   & =& \frac{lr'\klurd+\eta\,l'r\kldru}{\sqrt{|lr'|^2+|l'r|^2}},
	    \end{eqnarray}
	postselected with probability
	\begin{equation}
	\label{sloccprobpure}
	    P_\textrm{LR}=\prescript{}{D}{\braket{\Uppsi|\Pi_\textrm{LR}|\Uppsi}_D}=|lr'|^2+|l'r|^2.
	\end{equation}
	Notice that the two qubits in the final state $\ket{\Uppsi}_\textrm{LR}$ of Eq.\tilde\eqref{sloccstatepure} are distinguishable, since one of them is now localized in region $L$ while the other in region $R$.
	
	The first aspect that emerges from Eq.\tilde\eqref{sloccstatepure} is that the final state $\ket{\Uppsi}_\textrm{LR}$ is an entangled state, provided $l,l',r,r'\neq0$. Since the initial state was non-entangled, we thus conclude that \textit{the sLOCC protocol can be used to generate entanglement}\tilde\cite{slocc,experimentalslocc}. Remarkably, the superposition of states $\klurd$ and $\kldru$ is a direct consequence of the impossibility for the two detectors to understand which one of the two qubits they have detected, namely if the one with spin\tilde$\uparrow$ generated in $A$ or the one with spin $\downarrow$ generated in $B$. In other words, \textit{the origin of the quantum correlations} in the sLOCC-generated state of Eq.\tilde\eqref{sloccstatepure} is \textit{the no-which-way information} discussed in Section\tilde\ref{section1} deriving from the achieved spatial indistinguishability. For this reason, we say that deformations leading to indistinguishability \textit{activate} entanglement, while the sLOCC measurement allows to \textit{access} it. To further stress this point, we remark that $\ket{\Uppsi}_\textrm{LR}$ is non-entangled whenever at least one among $l,l',r,r'$ is null; indeed, this amounts to the scenario where (at least) one of the qubits is perfectly localized either on $L$ or on $R$, so that the coincidence click required by the sLOCC measurement allows to precisely track the origin of bothe the particles. This is the situation occurring, e.g., when no deformation is performed, so that $l=r'=1$ and $l'=r=0$: particles remain distinguishable and no entanglement is generated.
	
	From Eq.\tilde\eqref{indistgeneral} with $\vec{\alpha}=\{\emptyset\}$, $N=2$, and $S_1=L,\,S_2=R$, the amount of spatial indistinguishability obtained with the deformation can be properly quantified by the entropic measure introduced in Section\tilde\ref{sec:entropicmeasure}\tilde\cite{indistentanglprotection}
	\begin{equation}
	\label{indistinguishabilityslocc}
	    \mathcal{I}_\textrm{LR}=
	    -\dfrac{|l|^2\,|r'|^2}{\mathcal{Z}} \log_2 \dfrac{|l|^2\,|r'|^2}{\mathcal{Z}}
	    -\dfrac{|l'|^2\,|r|^2}{\mathcal{Z}}\log_2 \dfrac{|l'|^2\,|r|^2}{\mathcal{Z}},
	\end{equation}
	where $\mathcal{Z}=|l|^2\,|r'|^2+|l'|^2\,|r|^2$.
	Such a quantity takes into account the no-which-way information, taking the minimum value $\mathcal{I}=0$ when no overlap is present ($l=1,\,r'=1$ or $l'=1,\,r=1$: distinguishable particles) and the maximum one $\mathcal{I}=1$ when the overlap is maximum ($l=l'=r=r'=1/\sqrt{2}$: maximally indistinguishable particles).

The role of indistinguishability as a resource for quantum technologies within the sLOCC framework has been investigated by several recent experiments. Remarkably, in Ref.\tilde\cite{experimentalslocc} the authors have experimentally implemented the deformation+sLOCC protocol with two photons initially prepared in the state $\ket{\Uppsi}_\textrm{AB}$. They have performed quantum teleportation with the final state of Eq.\tilde\eqref{sloccstatepure}, thus showing that the achieved entanglement is physical. Furthermore, by directly accessing the value of $l,\,l',\,r,\,r'$ they fixed $l=r=1/\sqrt{2}$ to make $\mathcal{I}$ a function of just one parameter and showed that \textit{the amount of quantum correlations present in the state produced by the sLOCC protocol}, as quantified by the entanglement of formation\tilde\cite{concurrence}, \textit{is proportional to the degree of spatial indistinguishability achieved}. In particular, when $\mathcal{I}=1$ we see from Eq.\tilde\eqref{sloccstatepure} that the sLOCC process generates the maximally entangled state $\ket{\Uppsi}_\textrm{LR}^\textrm{max}=(\klurd+\eta\kldru)/\sqrt{2}$.
	
	In Refs.\tilde\cite{indistentanglprotection,Piccolini_2021,indistdynamicalprotection,piccolini2022indistinguishability} the authors considered the more realistic scenario where the deformation+sLOCC protocol is applied to two qubits in the presence of noise. Considering the maximally entangled Bell singlet state $\ket{1_-}:=(\kaubd-\kadbu)/\sqrt{2}$ as initial state, they analyzed the entanglement of formation of the system undergoing the detrimental action of two independent and localized noisy environments, both of the same type and modelled as phase damping channels, depolarizing channels, and amplitude damping channels. Noise has been considered acting before the deformation, when the particles are still distinguishable\tilde\cite{indistentanglprotection,Piccolini_2021}, after it, when the qubits are indistinguishable to the eyes of the environments\cite{indistdynamicalprotection}, and in both  situations altogether\tilde\cite{piccolini2022indistinguishability}. For general values of $l,\,l',\,r,\,r'$, the deformation is now found to generate a mixed state $\rd$ which the sLOCC projection transforms into (extending Eq.\tilde\eqref{sloccstatepure} to mixed states)
	\begin{equation}
	    \label{sloccstatemixed}
	    \rlr
	    =\frac{\hat{\Pi}_\textrm{LR}\,\rd\,\hat{\Pi}_\textrm{LR}}{\textrm{Tr}\left[\hat{\Pi}_\textrm{LR}\,\rd\right]},
	\end{equation}
	with postselection probability
	\begin{equation}
	\label{sloccprobmixed}
	    P_\textrm{LR}
	    =\textrm{Tr}\left[\hat{\Pi}_\textrm{LR}\,\rd\right].
	\end{equation}
	Remarkably, the process is found to lead to a restoration of the final quantum correlations present in $\rlr$ with respect to the ones characterizing the state immediately before the generation of the spatial indistinguishability. Furthermore, the amount of entanglement restored results to be proportional to the degree of spatial indistinguishability achieved. In particular, when $\mathcal{I}=1$ the deformation+sLOCC protocol is found to completely regenerate the maximally entangled initial state $\kom$ no matter how long the interaction time has been. Thus, \textit{the sLOCC operational framework can be used to restore}, partially or even completely, \textit{the amount of quantum correlations present in an initially entangled state spoiled by the detrimental interaction with noisy environments}.
	
	Another relevant element emerging from the sLOCC-prepared state, as can be noticed from Eq.\tilde\eqref{sloccstatepure}, is the factor $\eta:=e^{i\theta}$ encoding the exchange phase $\theta$, with $\theta=0$ for bosons and $\theta=\pi$ for fermions being at the core of the symmetrization postulate discussed in Section\tilde\ref{section2}.
	Although many decades have passed after the first formulation of the postulate, a first direct experimental measurement of the bosonic exchange phase has been only recently achieved with two photons in an all-optical setup\tilde\cite{tschernig2021direct}.
	This is mainly due to the difficulty in designing a setup \textit{manually} generating a superposition  between a reference state and its physically permuted one, from which later extrapolating the relative exchange phase via interferometry. Thanks to its reliance on spatial indistinguishability, the sLOCC process allows to avoid such a difficulty by letting $\theta$ naturally emerge. Exploiting this effect, in Refs.\tilde\cite{wang2022direct,lo2021directly} the authors designed and experimentally implemented an optical setup capable of directly measuring the exchange phase of two photons by applying interferometry to the sLOCC-produced state of Eq.\tilde\eqref{sloccstatepure}. Remarkably, the introduced theoretical setup is general and could be suitably adapted to directly measure the exchange phase of even fermions and anyons.
	
	Finally, spatial indistinguishability of identical particles undergoing the sLOCC measurement has been shown to provide a useful resource of quantum coherence yielding an advantage in quantum metrology\tilde\cite{Castellini2019metrology,sun2021experimental}, whereas the endurance of quantum coherence within systems of indistinguishable particles in non-dissipative noisy quantum networks was demonstrated in Ref.\tilde\cite{perez2018endurance}.
	
	It is interesting to highlight the connection between the deformation+sLOCC operational framework and the \textit{entanglement extraction} protocol\tilde\cite{killoran2014extracting}. In the latter, a single-mode state of indistinguishable particles is splitted over distinct modes. The resulting particle number distribution is then measured along such modes, postselecting only those states which respect a desired partition. Being the resulting modes distinct, this allows to access the entanglement between groups of identical particles whose accessibility was previously ruled out by their single-mode indistinguishability. In relation to this framework, the mode splitting operation is a particular case of deformation acting on already indistinguishable particles. Furthermore, deformations such as mode merging operations can be seen themselves as the preparation step required to achieve the entanglement extraction single-mode starting point. Furthermore, the particles distribution postselected measurement and the sLOCC projection are clearly related, since they both make quantum correlations accessible by making an indistinguishable state distinguishable. Nonetheless, while entanglement extraction focuses on the splitting of an already indistinguishable state to show that quantum correlations inaccessible within identical systems are actually physically meaningful and constitute useful resources in their own right\tilde\cite{killoran2014extracting,morrisPRX}, the sLOCC process presents itself as an alternative operational framework where indistinguishability is generated over previously arranged detection regions with the goal of generating, restoring, and/or manipulating entanglement in actual practical applications.

	\section{Conclusion}
	In conclusion, we have discussed and elucidated the distinction between the concepts of particle identity and particle indistinguishability in quantum mechanics.
	We have presented a concise review of the no-label approach as a suitable tool to deal with indistinguishable constituents, as introduced in Ref.\tilde\cite{nolabelappr} and further deepened in Refs\tilde\cite{compagno2018dealing,indistdynamicalprotection}.
	We have introduced a coherent formalization of deformations acting on either distinguishable or indistinguishable multipartite states, providing an extension of the indistinguishability entropic measure introduced in Ref.\tilde\cite{indistdynamicalprotection} to the general $N$-partite scenario.
	We have highlighted the relevance of deformations as operations exploitable to activate quantum correlations to be later accessed within the sLOCC operational framework.
	Finally, we have briefly discussed the relations between the sLOCC protocol and the entanglement extraction one as operational frameworks.
	
	Given the results presented in this work, we believe that deformations, together with the sLOCC operational framework, have the potential to become a useful technique for many real-world applications exploiting quantum technologies. Indeed, identical particles constitute the main building blocks of platforms such as quantum networks, quantum computers, and quantum measurement systems. For instance, spatial indistinguishability of identical constituents generated by properly tuned deformations could be exploited to shield from noise the fundamental quantum correlation properties required for quantum cryptographic protocols, or the coherence of qubits used to run quantum algorithms. Furthermore, the entanglement-restoration characteristics of the presented techniques could be further investigated to preserve the super-sensitivity of states carrying information in quantum sensing and metrology protocols.

\textbf{Acknowledgments.} M.P., F.N, and R.L.F. acknowledge Alessia Castellini and Giuseppe Compagno for fruitful discussions and feedback. R.M. thanks support from NSERC, MEI and the CRC program in Canada. R.L.F. acknowledges support from Unione Europea -- NextGenerationEU -- fondi MUR D.M. 737/2021 -- progetto di ricerca ``IRISQ''.

    \enlargethispage{20pt}









\begin{thebibliography}{36}%
\makeatletter
\providecommand \@ifxundefined [1]{%
 \@ifx{#1\undefined}
}%
\providecommand \@ifnum [1]{%
 \ifnum #1\expandafter \@firstoftwo
 \else \expandafter \@secondoftwo
 \fi
}%
\providecommand \@ifx [1]{%
 \ifx #1\expandafter \@firstoftwo
 \else \expandafter \@secondoftwo
 \fi
}%
\providecommand \natexlab [1]{#1}%
\providecommand \enquote  [1]{``#1''}%
\providecommand \bibnamefont  [1]{#1}%
\providecommand \bibfnamefont [1]{#1}%
\providecommand \citenamefont [1]{#1}%
\providecommand \href@noop [0]{\@secondoftwo}%
\providecommand \href [0]{\begingroup \@sanitize@url \@href}%
\providecommand \@href[1]{\@@startlink{#1}\@@href}%
\providecommand \@@href[1]{\endgroup#1\@@endlink}%
\providecommand \@sanitize@url [0]{\catcode `\\12\catcode `\$12\catcode
  `\&12\catcode `\#12\catcode `\^12\catcode `\_12\catcode `\%12\relax}%
\providecommand \@@startlink[1]{}%
\providecommand \@@endlink[0]{}%
\providecommand \url  [0]{\begingroup\@sanitize@url \@url }%
\providecommand \@url [1]{\endgroup\@href {#1}{\urlprefix }}%
\providecommand \urlprefix  [0]{URL }%
\providecommand \Eprint [0]{\href }%
\providecommand \doibase [0]{https://doi.org/}%
\providecommand \selectlanguage [0]{\@gobble}%
\providecommand \bibinfo  [0]{\@secondoftwo}%
\providecommand \bibfield  [0]{\@secondoftwo}%
\providecommand \translation [1]{[#1]}%
\providecommand \BibitemOpen [0]{}%
\providecommand \bibitemStop [0]{}%
\providecommand \bibitemNoStop [0]{.\EOS\space}%
\providecommand \EOS [0]{\spacefactor3000\relax}%
\providecommand \BibitemShut  [1]{\csname bibitem#1\endcsname}%
\let\auto@bib@innerbib\@empty
\bibitem [{\citenamefont {Cohen-Tannoudji}\ \emph {et~al.}(2006)\citenamefont
  {Cohen-Tannoudji}, \citenamefont {Diu}, \citenamefont {Laloe},\ and\
  \citenamefont {Dui}}]{cohen2006quantum}%
  \BibitemOpen
  \bibfield  {author} {\bibinfo {author} {\bibfnamefont {C.}~\bibnamefont
  {Cohen-Tannoudji}}, \bibinfo {author} {\bibfnamefont {B.}~\bibnamefont
  {Diu}}, \bibinfo {author} {\bibfnamefont {F.}~\bibnamefont {Laloe}},\ and\
  \bibinfo {author} {\bibfnamefont {B.}~\bibnamefont {Dui}},\ }\href@noop {}
  {\emph {\bibinfo {title} {Quantum Mechanics (2 vol. set)}}}\ (\bibinfo
  {publisher} {Wiley-Interscience},\ \bibinfo {year} {2006})\BibitemShut
  {NoStop}%
\bibitem [{\citenamefont {Ghirardi}\ \emph {et~al.}(2002)\citenamefont
  {Ghirardi}, \citenamefont {Marinatto},\ and\ \citenamefont
  {Weber}}]{ghirardi2002entanglement}%
  \BibitemOpen
  \bibfield  {author} {\bibinfo {author} {\bibfnamefont {G.}~\bibnamefont
  {Ghirardi}}, \bibinfo {author} {\bibfnamefont {L.}~\bibnamefont
  {Marinatto}},\ and\ \bibinfo {author} {\bibfnamefont {T.}~\bibnamefont
  {Weber}},\ }\bibfield  {title} {\bibinfo {title} {Entanglement and properties
  of composite quantum systems: a conceptual and mathematical analysis},\
  }\href@noop {} {\bibfield  {journal} {\bibinfo  {journal} {J. Stat. Phys.}\
  }\textbf {\bibinfo {volume} {108}},\ \bibinfo {pages} {49} (\bibinfo {year}
  {2002})}\BibitemShut {NoStop}%
\bibitem [{\citenamefont {Tichy}\ \emph {et~al.}(2013)\citenamefont {Tichy},
  \citenamefont {{de Melo}}, \citenamefont {Kus}, \citenamefont {Mintert},\
  and\ \citenamefont {Buchleitner}}]{tichyFort}%
  \BibitemOpen
  \bibfield  {author} {\bibinfo {author} {\bibfnamefont {M.~C.}\ \bibnamefont
  {Tichy}}, \bibinfo {author} {\bibfnamefont {F.}~\bibnamefont {{de Melo}}},
  \bibinfo {author} {\bibfnamefont {M.}~\bibnamefont {Kus}}, \bibinfo {author}
  {\bibfnamefont {F.}~\bibnamefont {Mintert}},\ and\ \bibinfo {author}
  {\bibfnamefont {A.}~\bibnamefont {Buchleitner}},\ }\bibfield  {title}
  {\bibinfo {title} {Entanglement of identical particles and the detection
  process},\ }\href@noop {} {\bibfield  {journal} {\bibinfo  {journal}
  {Fortschr. Phys.}\ }\textbf {\bibinfo {volume} {61}},\ \bibinfo {pages} {225}
  (\bibinfo {year} {2013})}\BibitemShut {NoStop}%
\bibitem [{\citenamefont {Lo~Franco}\ and\ \citenamefont
  {Compagno}(2016)}]{nolabelappr}%
  \BibitemOpen
  \bibfield  {author} {\bibinfo {author} {\bibfnamefont {R.}~\bibnamefont
  {Lo~Franco}}\ and\ \bibinfo {author} {\bibfnamefont {G.}~\bibnamefont
  {Compagno}},\ }\bibfield  {title} {\bibinfo {title} {Quantum entanglement of
  identical particles by standard information-theoretic notions},\ }\href@noop
  {} {\bibfield  {journal} {\bibinfo  {journal} {Sci. Rep.}\ }\textbf {\bibinfo
  {volume} {6}},\ \bibinfo {pages} {20603} (\bibinfo {year}
  {2016})}\BibitemShut {NoStop}%
\bibitem [{\citenamefont {Herbut}\ and\ \citenamefont
  {Vujicic}(1987)}]{Herbut_1987}%
  \BibitemOpen
  \bibfield  {author} {\bibinfo {author} {\bibfnamefont {F.}~\bibnamefont
  {Herbut}}\ and\ \bibinfo {author} {\bibfnamefont {M.}~\bibnamefont
  {Vujicic}},\ }\bibfield  {title} {\bibinfo {title} {Irrelevance of the pauli
  principle in distant correlations between identical fermions},\ }\href
  {https://doi.org/10.1088/0305-4470/20/16/030} {\bibfield  {journal} {\bibinfo
   {journal} {J. Phys. A: Math. Gen.}\ }\textbf {\bibinfo {volume} {20}},\
  \bibinfo {pages} {5555} (\bibinfo {year} {1987})}\BibitemShut {NoStop}%
\bibitem [{\citenamefont {Tichy}\ \emph {et~al.}(2011)\citenamefont {Tichy},
  \citenamefont {Mintert},\ and\ \citenamefont {Buchleitner}}]{tichy}%
  \BibitemOpen
  \bibfield  {author} {\bibinfo {author} {\bibfnamefont {M.~C.}\ \bibnamefont
  {Tichy}}, \bibinfo {author} {\bibfnamefont {F.}~\bibnamefont {Mintert}},\
  and\ \bibinfo {author} {\bibfnamefont {A.}~\bibnamefont {Buchleitner}},\
  }\bibfield  {title} {\bibinfo {title} {Essential entanglement for atomic and
  molecular physics},\ }\href@noop {} {\bibfield  {journal} {\bibinfo
  {journal} {J. Phys. B: At. Mol. Opt. Phys.}\ }\textbf {\bibinfo {volume}
  {44}},\ \bibinfo {pages} {192001} (\bibinfo {year} {2011})}\BibitemShut
  {NoStop}%
\bibitem [{\citenamefont {Peres}(2002)}]{Peresbook}%
  \BibitemOpen
  \bibfield  {author} {\bibinfo {author} {\bibfnamefont {A.}~\bibnamefont
  {Peres}},\ }\href@noop {} {\emph {\bibinfo {title} {Quantum Theory: Concepts
  and Methods}}}\ (\bibinfo  {publisher} {Springer},\ \bibinfo {year}
  {2002})\BibitemShut {NoStop}%
\bibitem [{\citenamefont {Nosrati}\ \emph
  {et~al.}(2020{\natexlab{a}})\citenamefont {Nosrati}, \citenamefont
  {Castellini}, \citenamefont {Compagno},\ and\ \citenamefont
  {Lo~Franco}}]{indistdynamicalprotection}%
  \BibitemOpen
  \bibfield  {author} {\bibinfo {author} {\bibfnamefont {F.}~\bibnamefont
  {Nosrati}}, \bibinfo {author} {\bibfnamefont {A.}~\bibnamefont {Castellini}},
  \bibinfo {author} {\bibfnamefont {G.}~\bibnamefont {Compagno}},\ and\
  \bibinfo {author} {\bibfnamefont {R.}~\bibnamefont {Lo~Franco}},\ }\bibfield
  {title} {\bibinfo {title} {Dynamics of spatially indistinguishable particles
  and quantum entanglement protection},\ }\href@noop {} {\bibfield  {journal}
  {\bibinfo  {journal} {Phys. Rev. A}\ }\textbf {\bibinfo {volume} {102}},\
  \bibinfo {pages} {062429} (\bibinfo {year} {2020}{\natexlab{a}})}\BibitemShut
  {NoStop}%
\bibitem [{\citenamefont {Piccolini}\ \emph {et~al.}(2021)\citenamefont
  {Piccolini}, \citenamefont {Nosrati}, \citenamefont {Compagno}, \citenamefont
  {Livreri}, \citenamefont {Morandotti},\ and\ \citenamefont
  {Lo~Franco}}]{Piccolini_2021}%
  \BibitemOpen
  \bibfield  {author} {\bibinfo {author} {\bibfnamefont {M.}~\bibnamefont
  {Piccolini}}, \bibinfo {author} {\bibfnamefont {F.}~\bibnamefont {Nosrati}},
  \bibinfo {author} {\bibfnamefont {G.}~\bibnamefont {Compagno}}, \bibinfo
  {author} {\bibfnamefont {P.}~\bibnamefont {Livreri}}, \bibinfo {author}
  {\bibfnamefont {R.}~\bibnamefont {Morandotti}},\ and\ \bibinfo {author}
  {\bibfnamefont {R.}~\bibnamefont {Lo~Franco}},\ }\bibfield  {title} {\bibinfo
  {title} {Entanglement robustness via spatial deformation of identical
  particle wave functions},\ }\href@noop {} {\bibfield  {journal} {\bibinfo
  {journal} {Entropy}\ }\textbf {\bibinfo {volume} {23}},\ \bibinfo {pages}
  {708} (\bibinfo {year} {2021})}\BibitemShut {NoStop}%
\bibitem [{\citenamefont {Piccolini}\ \emph {et~al.}(2022)\citenamefont
  {Piccolini}, \citenamefont {Nosrati}, \citenamefont {Morandotti},\ and\
  \citenamefont {Lo~Franco}}]{piccolini2022indistinguishability}%
  \BibitemOpen
  \bibfield  {author} {\bibinfo {author} {\bibfnamefont {M.}~\bibnamefont
  {Piccolini}}, \bibinfo {author} {\bibfnamefont {F.}~\bibnamefont {Nosrati}},
  \bibinfo {author} {\bibfnamefont {R.}~\bibnamefont {Morandotti}},\ and\
  \bibinfo {author} {\bibfnamefont {R.}~\bibnamefont {Lo~Franco}},\ }\bibfield
  {title} {\bibinfo {title} {Indistinguishability-enhanced entanglement
  recovery by spatially localized operations and classical communication},\
  }\href@noop {} {\bibfield  {journal} {\bibinfo  {journal} {Open Sys. Info.
  Dyn., in press. ArXiv preprint arXiv:2201.13365 [quant-ph]}\ } (\bibinfo
  {year} {2022})}\BibitemShut {NoStop}%
\bibitem [{\citenamefont {Compagno}\ \emph {et~al.}(2018)\citenamefont
  {Compagno}, \citenamefont {Castellini},\ and\ \citenamefont
  {Lo~Franco}}]{compagno2018dealing}%
  \BibitemOpen
  \bibfield  {author} {\bibinfo {author} {\bibfnamefont {G.}~\bibnamefont
  {Compagno}}, \bibinfo {author} {\bibfnamefont {A.}~\bibnamefont
  {Castellini}},\ and\ \bibinfo {author} {\bibfnamefont {R.}~\bibnamefont
  {Lo~Franco}},\ }\bibfield  {title} {\bibinfo {title} {Dealing with
  indistinguishable particles and their entanglement},\ }\href@noop {}
  {\bibfield  {journal} {\bibinfo  {journal} {Phil. Trans. R. Soc. A}\ }\textbf
  {\bibinfo {volume} {376}},\ \bibinfo {pages} {20170317} (\bibinfo {year}
  {2018})}\BibitemShut {NoStop}%
\bibitem [{\citenamefont {Nosrati}\ \emph
  {et~al.}(2020{\natexlab{b}})\citenamefont {Nosrati}, \citenamefont
  {Castellini}, \citenamefont {Compagno},\ and\ \citenamefont
  {Lo~Franco}}]{indistentanglprotection}%
  \BibitemOpen
  \bibfield  {author} {\bibinfo {author} {\bibfnamefont {F.}~\bibnamefont
  {Nosrati}}, \bibinfo {author} {\bibfnamefont {A.}~\bibnamefont {Castellini}},
  \bibinfo {author} {\bibfnamefont {G.}~\bibnamefont {Compagno}},\ and\
  \bibinfo {author} {\bibfnamefont {R.}~\bibnamefont {Lo~Franco}},\ }\bibfield
  {title} {\bibinfo {title} {Robust entanglement preparation against noise by
  controlling spatial indistinguishability},\ }\href@noop {} {\bibfield
  {journal} {\bibinfo  {journal} {npj Quant. Inf.}\ }\textbf {\bibinfo {volume}
  {6}},\ \bibinfo {pages} {1} (\bibinfo {year}
  {2020}{\natexlab{b}})}\BibitemShut {NoStop}%
\bibitem [{\citenamefont {Ghirardi}\ and\ \citenamefont
  {Marinatto}(2004)}]{ghirardi}%
  \BibitemOpen
  \bibfield  {author} {\bibinfo {author} {\bibfnamefont {G.}~\bibnamefont
  {Ghirardi}}\ and\ \bibinfo {author} {\bibfnamefont {L.}~\bibnamefont
  {Marinatto}},\ }\bibfield  {title} {\bibinfo {title} {General criterion for
  the entanglement of two indistinguishable particles},\ }\href@noop {}
  {\bibfield  {journal} {\bibinfo  {journal} {Phys. Rev. A}\ }\textbf {\bibinfo
  {volume} {70}},\ \bibinfo {pages} {012109} (\bibinfo {year}
  {2004})}\BibitemShut {NoStop}%
\bibitem [{\citenamefont {Li}\ \emph {et~al.}(2001)\citenamefont {Li},
  \citenamefont {Zeng}, \citenamefont {Liu},\ and\ \citenamefont
  {Long}}]{Li2001PRA}%
  \BibitemOpen
  \bibfield  {author} {\bibinfo {author} {\bibfnamefont {Y.-S.}\ \bibnamefont
  {Li}}, \bibinfo {author} {\bibfnamefont {B.}~\bibnamefont {Zeng}}, \bibinfo
  {author} {\bibfnamefont {X.-S.}\ \bibnamefont {Liu}},\ and\ \bibinfo {author}
  {\bibfnamefont {G.-L.}\ \bibnamefont {Long}},\ }\bibfield  {title} {\bibinfo
  {title} {Entanglement in a {two-identical-particle} system},\ }\href@noop {}
  {\bibfield  {journal} {\bibinfo  {journal} {Phys. Rev. A}\ }\textbf {\bibinfo
  {volume} {64}},\ \bibinfo {pages} {054302} (\bibinfo {year}
  {2001})}\BibitemShut {NoStop}%
\bibitem [{\citenamefont {Paskauskas}\ and\ \citenamefont
  {You}(2001)}]{Paskauskas2001PRA}%
  \BibitemOpen
  \bibfield  {author} {\bibinfo {author} {\bibfnamefont {R.}~\bibnamefont
  {Paskauskas}}\ and\ \bibinfo {author} {\bibfnamefont {L.}~\bibnamefont
  {You}},\ }\bibfield  {title} {\bibinfo {title} {Quantum correlations in
  {two-boson} wave functions},\ }\href@noop {} {\bibfield  {journal} {\bibinfo
  {journal} {Phys. Rev. A}\ }\textbf {\bibinfo {volume} {64}},\ \bibinfo
  {pages} {042310} (\bibinfo {year} {2001})}\BibitemShut {NoStop}%
\bibitem [{\citenamefont {Schliemann}\ \emph {et~al.}(2001)\citenamefont
  {Schliemann}, \citenamefont {Cirac}, \citenamefont {Kus}, \citenamefont
  {Lewenstein},\ and\ \citenamefont {Loss}}]{cirac2001PRA}%
  \BibitemOpen
  \bibfield  {author} {\bibinfo {author} {\bibfnamefont {J.}~\bibnamefont
  {Schliemann}}, \bibinfo {author} {\bibfnamefont {J.~I.}\ \bibnamefont
  {Cirac}}, \bibinfo {author} {\bibfnamefont {M.}~\bibnamefont {Kus}}, \bibinfo
  {author} {\bibfnamefont {M.}~\bibnamefont {Lewenstein}},\ and\ \bibinfo
  {author} {\bibfnamefont {D.}~\bibnamefont {Loss}},\ }\bibfield  {title}
  {\bibinfo {title} {Quantum correlations in two-fermion systems},\ }\href@noop
  {} {\bibfield  {journal} {\bibinfo  {journal} {Phys. Rev. A}\ }\textbf
  {\bibinfo {volume} {64}},\ \bibinfo {pages} {022303} (\bibinfo {year}
  {2001})}\BibitemShut {NoStop}%
\bibitem [{\citenamefont {Zanardi}(2002)}]{zanardiPRA}%
  \BibitemOpen
  \bibfield  {author} {\bibinfo {author} {\bibfnamefont {P.}~\bibnamefont
  {Zanardi}},\ }\bibfield  {title} {\bibinfo {title} {Quantum entanglement in
  fermionic lattices},\ }\href@noop {} {\bibfield  {journal} {\bibinfo
  {journal} {Phys. Rev. A}\ }\textbf {\bibinfo {volume} {65}},\ \bibinfo
  {pages} {042101} (\bibinfo {year} {2002})}\BibitemShut {NoStop}%
\bibitem [{\citenamefont {Morris}\ \emph {et~al.}(2020)\citenamefont {Morris},
  \citenamefont {Yadin}, \citenamefont {Fadel}, \citenamefont {Zibold},
  \citenamefont {Treutlein},\ and\ \citenamefont {Adesso}}]{morrisPRX}%
  \BibitemOpen
  \bibfield  {author} {\bibinfo {author} {\bibfnamefont {B.}~\bibnamefont
  {Morris}}, \bibinfo {author} {\bibfnamefont {B.}~\bibnamefont {Yadin}},
  \bibinfo {author} {\bibfnamefont {M.}~\bibnamefont {Fadel}}, \bibinfo
  {author} {\bibfnamefont {T.}~\bibnamefont {Zibold}}, \bibinfo {author}
  {\bibfnamefont {P.}~\bibnamefont {Treutlein}},\ and\ \bibinfo {author}
  {\bibfnamefont {G.}~\bibnamefont {Adesso}},\ }\bibfield  {title} {\bibinfo
  {title} {Entanglement between identical particles is a useful and consistent
  resource},\ }\href@noop {} {\bibfield  {journal} {\bibinfo  {journal} {Phys.
  Rev. X}\ }\textbf {\bibinfo {volume} {10}},\ \bibinfo {pages} {041012}
  (\bibinfo {year} {2020})}\BibitemShut {NoStop}%
\bibitem [{\citenamefont {Eckert}\ \emph {et~al.}(2002)\citenamefont {Eckert},
  \citenamefont {Schliemann}, \citenamefont {Bruss},\ and\ \citenamefont
  {Lewenstein}}]{eckert2002AnnPhys}%
  \BibitemOpen
  \bibfield  {author} {\bibinfo {author} {\bibfnamefont {K.}~\bibnamefont
  {Eckert}}, \bibinfo {author} {\bibfnamefont {J.}~\bibnamefont {Schliemann}},
  \bibinfo {author} {\bibfnamefont {D.}~\bibnamefont {Bruss}},\ and\ \bibinfo
  {author} {\bibfnamefont {M.}~\bibnamefont {Lewenstein}},\ }\bibfield  {title}
  {\bibinfo {title} {Quantum correlations in systems of indistinguishable
  particles},\ }\href@noop {} {\bibfield  {journal} {\bibinfo  {journal} {Ann.
  Phys.}\ }\textbf {\bibinfo {volume} {299}},\ \bibinfo {pages} {88} (\bibinfo
  {year} {2002})}\BibitemShut {NoStop}%
\bibitem [{\citenamefont {Balachandran}\ \emph {et~al.}(2013)\citenamefont
  {Balachandran}, \citenamefont {Govindarajan}, \citenamefont {de~Queiroz},\
  and\ \citenamefont {Reyes-Lega}}]{balachandranPRL}%
  \BibitemOpen
  \bibfield  {author} {\bibinfo {author} {\bibfnamefont {A.~P.}\ \bibnamefont
  {Balachandran}}, \bibinfo {author} {\bibfnamefont {T.~R.}\ \bibnamefont
  {Govindarajan}}, \bibinfo {author} {\bibfnamefont {A.~R.}\ \bibnamefont
  {de~Queiroz}},\ and\ \bibinfo {author} {\bibfnamefont {A.~F.}\ \bibnamefont
  {Reyes-Lega}},\ }\bibfield  {title} {\bibinfo {title} {Entanglement and
  particle identity: A unifying approach},\ }\href@noop {} {\bibfield
  {journal} {\bibinfo  {journal} {Phys. Rev. Lett.}\ }\textbf {\bibinfo
  {volume} {110}},\ \bibinfo {pages} {080503} (\bibinfo {year}
  {2013})}\BibitemShut {NoStop}%
\bibitem [{\citenamefont {Cunden}\ \emph {et~al.}(2014)\citenamefont {Cunden},
  \citenamefont {{Di Martino}}, \citenamefont {Facchi},\ and\ \citenamefont
  {Florio}}]{facchiIJQI}%
  \BibitemOpen
  \bibfield  {author} {\bibinfo {author} {\bibfnamefont {F.~D.}\ \bibnamefont
  {Cunden}}, \bibinfo {author} {\bibfnamefont {S.}~\bibnamefont {{Di
  Martino}}}, \bibinfo {author} {\bibfnamefont {P.}~\bibnamefont {Facchi}},\
  and\ \bibinfo {author} {\bibfnamefont {G.}~\bibnamefont {Florio}},\
  }\bibfield  {title} {\bibinfo {title} {Spatial separation and entanglement of
  identical particles},\ }\href@noop {} {\bibfield  {journal} {\bibinfo
  {journal} {Int. J. Quantum Inform.}\ }\textbf {\bibinfo {volume} {12}},\
  \bibinfo {pages} {461001} (\bibinfo {year} {2014})}\BibitemShut {NoStop}%
\bibitem [{\citenamefont {Sasaki}\ \emph {et~al.}(2011)\citenamefont {Sasaki},
  \citenamefont {Ichikawa},\ and\ \citenamefont {Tsutsui}}]{sasaki2011PRA}%
  \BibitemOpen
  \bibfield  {author} {\bibinfo {author} {\bibfnamefont {T.}~\bibnamefont
  {Sasaki}}, \bibinfo {author} {\bibfnamefont {T.}~\bibnamefont {Ichikawa}},\
  and\ \bibinfo {author} {\bibfnamefont {I.}~\bibnamefont {Tsutsui}},\
  }\bibfield  {title} {\bibinfo {title} {Entanglement of indistinguishable
  particles},\ }\href@noop {} {\bibfield  {journal} {\bibinfo  {journal} {Phys.
  Rev. A}\ }\textbf {\bibinfo {volume} {83}},\ \bibinfo {pages} {012113}
  (\bibinfo {year} {2011})}\BibitemShut {NoStop}%
\bibitem [{\citenamefont {Bose}\ and\ \citenamefont
  {Home}(2002)}]{bose2002indisting}%
  \BibitemOpen
  \bibfield  {author} {\bibinfo {author} {\bibfnamefont {S.}~\bibnamefont
  {Bose}}\ and\ \bibinfo {author} {\bibfnamefont {D.}~\bibnamefont {Home}},\
  }\bibfield  {title} {\bibinfo {title} {Generic entanglement generation,
  quantum statistics, and complementarity},\ }\href@noop {} {\bibfield
  {journal} {\bibinfo  {journal} {Phys. Rev. Lett.}\ }\textbf {\bibinfo
  {volume} {88}},\ \bibinfo {pages} {050401} (\bibinfo {year}
  {2002})}\BibitemShut {NoStop}%
\bibitem [{\citenamefont {Bose}\ and\ \citenamefont {Home}(2013)}]{bose2013}%
  \BibitemOpen
  \bibfield  {author} {\bibinfo {author} {\bibfnamefont {S.}~\bibnamefont
  {Bose}}\ and\ \bibinfo {author} {\bibfnamefont {D.}~\bibnamefont {Home}},\
  }\bibfield  {title} {\bibinfo {title} {Duality in entanglement enabling a
  test of quantum indistinguishability unaffected by interactions},\
  }\href@noop {} {\bibfield  {journal} {\bibinfo  {journal} {Phys. Rev. Lett.}\
  }\textbf {\bibinfo {volume} {110}},\ \bibinfo {pages} {140404} (\bibinfo
  {year} {2013})}\BibitemShut {NoStop}%
\bibitem [{\citenamefont {Killoran}\ \emph {et~al.}(2014)\citenamefont
  {Killoran}, \citenamefont {Cramer},\ and\ \citenamefont
  {Plenio}}]{killoran2014extracting}%
  \BibitemOpen
  \bibfield  {author} {\bibinfo {author} {\bibfnamefont {N.}~\bibnamefont
  {Killoran}}, \bibinfo {author} {\bibfnamefont {M.}~\bibnamefont {Cramer}},\
  and\ \bibinfo {author} {\bibfnamefont {M.~B.}\ \bibnamefont {Plenio}},\
  }\bibfield  {title} {\bibinfo {title} {Extracting entanglement from identical
  particles},\ }\href@noop {} {\bibfield  {journal} {\bibinfo  {journal} {Phys.
  Rev. Lett.}\ }\textbf {\bibinfo {volume} {112}},\ \bibinfo {pages} {150501}
  (\bibinfo {year} {2014})}\BibitemShut {NoStop}%
\bibitem [{\citenamefont {Sciara}\ \emph {et~al.}(2017)\citenamefont {Sciara},
  \citenamefont {{Lo Franco}},\ and\ \citenamefont {Compagno}}]{sciaraSchmidt}%
  \BibitemOpen
  \bibfield  {author} {\bibinfo {author} {\bibfnamefont {S.}~\bibnamefont
  {Sciara}}, \bibinfo {author} {\bibfnamefont {R.}~\bibnamefont {{Lo
  Franco}}},\ and\ \bibinfo {author} {\bibfnamefont {G.}~\bibnamefont
  {Compagno}},\ }\bibfield  {title} {\bibinfo {title} {Universality of
  {Schmidt} decomposition and particle identity},\ }\href@noop {} {\bibfield
  {journal} {\bibinfo  {journal} {Sci. Rep.}\ }\textbf {\bibinfo {volume}
  {7}},\ \bibinfo {pages} {44675} (\bibinfo {year} {2017})}\BibitemShut
  {NoStop}%
\bibitem [{\citenamefont {Lo~Franco}\ and\ \citenamefont
  {Compagno}(2018)}]{slocc}%
  \BibitemOpen
  \bibfield  {author} {\bibinfo {author} {\bibfnamefont {R.}~\bibnamefont
  {Lo~Franco}}\ and\ \bibinfo {author} {\bibfnamefont {G.}~\bibnamefont
  {Compagno}},\ }\bibfield  {title} {\bibinfo {title} {Indistinguishability of
  elementary systems as a resource for quantum information processing},\
  }\href@noop {} {\bibfield  {journal} {\bibinfo  {journal} {Phys. Rev. Lett.}\
  }\textbf {\bibinfo {volume} {120}},\ \bibinfo {pages} {240403} (\bibinfo
  {year} {2018})}\BibitemShut {NoStop}%
\bibitem [{\citenamefont {Horodecki}\ \emph {et~al.}(2009)\citenamefont
  {Horodecki}, \citenamefont {Horodecki}, \citenamefont {Horodecki},\ and\
  \citenamefont {Horodecki}}]{horodecki2009quantum}%
  \BibitemOpen
  \bibfield  {author} {\bibinfo {author} {\bibfnamefont {R.}~\bibnamefont
  {Horodecki}}, \bibinfo {author} {\bibfnamefont {P.}~\bibnamefont
  {Horodecki}}, \bibinfo {author} {\bibfnamefont {M.}~\bibnamefont
  {Horodecki}},\ and\ \bibinfo {author} {\bibfnamefont {K.}~\bibnamefont
  {Horodecki}},\ }\bibfield  {title} {\bibinfo {title} {Quantum entanglement},\
  }\href@noop {} {\bibfield  {journal} {\bibinfo  {journal} {Rev. Mod. Phys.}\
  }\textbf {\bibinfo {volume} {81}},\ \bibinfo {pages} {865} (\bibinfo {year}
  {2009})}\BibitemShut {NoStop}%
\bibitem [{\citenamefont {Sun}\ \emph {et~al.}(2020)\citenamefont {Sun},
  \citenamefont {Wang}, \citenamefont {Liu}, \citenamefont {Xu}, \citenamefont
  {Xu}, \citenamefont {Li}, \citenamefont {Guo}, \citenamefont {Castellini},
  \citenamefont {Nosrati}, \citenamefont {Compagno},\ and\ \citenamefont
  {Lo~Franco}}]{experimentalslocc}%
  \BibitemOpen
  \bibfield  {author} {\bibinfo {author} {\bibfnamefont {K.}~\bibnamefont
  {Sun}}, \bibinfo {author} {\bibfnamefont {Y.}~\bibnamefont {Wang}}, \bibinfo
  {author} {\bibfnamefont {Z.-H.}\ \bibnamefont {Liu}}, \bibinfo {author}
  {\bibfnamefont {X.-Y.}\ \bibnamefont {Xu}}, \bibinfo {author} {\bibfnamefont
  {J.-S.}\ \bibnamefont {Xu}}, \bibinfo {author} {\bibfnamefont {C.-F.}\
  \bibnamefont {Li}}, \bibinfo {author} {\bibfnamefont {G.-C.}\ \bibnamefont
  {Guo}}, \bibinfo {author} {\bibfnamefont {A.}~\bibnamefont {Castellini}},
  \bibinfo {author} {\bibfnamefont {F.}~\bibnamefont {Nosrati}}, \bibinfo
  {author} {\bibfnamefont {G.}~\bibnamefont {Compagno}},\ and\ \bibinfo
  {author} {\bibfnamefont {R.}~\bibnamefont {Lo~Franco}},\ }\bibfield  {title}
  {\bibinfo {title} {Experimental quantum entanglement and teleportation by
  tuning remote spatial indistinguishability of independent photons},\
  }\href@noop {} {\bibfield  {journal} {\bibinfo  {journal} {Opt. Lett.}\
  }\textbf {\bibinfo {volume} {45}},\ \bibinfo {pages} {6410} (\bibinfo {year}
  {2020})}\BibitemShut {NoStop}%
\bibitem [{\citenamefont {Wootters}(1998)}]{concurrence}%
  \BibitemOpen
  \bibfield  {author} {\bibinfo {author} {\bibfnamefont {W.~K.}\ \bibnamefont
  {Wootters}},\ }\bibfield  {title} {\bibinfo {title} {Entanglement of
  formation of an arbitrary state of two qubits},\ }\href@noop {} {\bibfield
  {journal} {\bibinfo  {journal} {Phys. Rev. Lett.}\ }\textbf {\bibinfo
  {volume} {80}},\ \bibinfo {pages} {2245–2248} (\bibinfo {year}
  {1998})}\BibitemShut {NoStop}%
\bibitem [{\citenamefont {Tschernig}\ \emph {et~al.}(2021)\citenamefont
  {Tschernig}, \citenamefont {M{\"u}ller}, \citenamefont {Smoor}, \citenamefont
  {Kroh}, \citenamefont {Wolters}, \citenamefont {Benson}, \citenamefont
  {Busch},\ and\ \citenamefont {P{\'e}rez-Leija}}]{tschernig2021direct}%
  \BibitemOpen
  \bibfield  {author} {\bibinfo {author} {\bibfnamefont {K.}~\bibnamefont
  {Tschernig}}, \bibinfo {author} {\bibfnamefont {C.}~\bibnamefont
  {M{\"u}ller}}, \bibinfo {author} {\bibfnamefont {M.}~\bibnamefont {Smoor}},
  \bibinfo {author} {\bibfnamefont {T.}~\bibnamefont {Kroh}}, \bibinfo {author}
  {\bibfnamefont {J.}~\bibnamefont {Wolters}}, \bibinfo {author} {\bibfnamefont
  {O.}~\bibnamefont {Benson}}, \bibinfo {author} {\bibfnamefont
  {K.}~\bibnamefont {Busch}},\ and\ \bibinfo {author} {\bibfnamefont
  {A.}~\bibnamefont {P{\'e}rez-Leija}},\ }\bibfield  {title} {\bibinfo {title}
  {Direct observation of the particle exchange phase of photons},\ }\href@noop
  {} {\bibfield  {journal} {\bibinfo  {journal} {Nat. Photon.}\ }\textbf
  {\bibinfo {volume} {15}},\ \bibinfo {pages} {671} (\bibinfo {year}
  {2021})}\BibitemShut {NoStop}%
\bibitem [{\citenamefont {Wang}\ \emph {et~al.}(2022)\citenamefont {Wang},
  \citenamefont {Piccolini}, \citenamefont {Hao}, \citenamefont {Liu},
  \citenamefont {Sun}, \citenamefont {Xu}, \citenamefont {Li}, \citenamefont
  {Guo}, \citenamefont {Morandotti}, \citenamefont {Compagno} \emph
  {et~al.}}]{wang2022direct}%
  \BibitemOpen
  \bibfield  {author} {\bibinfo {author} {\bibfnamefont {Y.}~\bibnamefont
  {Wang}}, \bibinfo {author} {\bibfnamefont {M.}~\bibnamefont {Piccolini}},
  \bibinfo {author} {\bibfnamefont {Z.-Y.}\ \bibnamefont {Hao}}, \bibinfo
  {author} {\bibfnamefont {Z.-H.}\ \bibnamefont {Liu}}, \bibinfo {author}
  {\bibfnamefont {K.}~\bibnamefont {Sun}}, \bibinfo {author} {\bibfnamefont
  {J.-S.}\ \bibnamefont {Xu}}, \bibinfo {author} {\bibfnamefont {C.-F.}\
  \bibnamefont {Li}}, \bibinfo {author} {\bibfnamefont {G.-C.}\ \bibnamefont
  {Guo}}, \bibinfo {author} {\bibfnamefont {R.}~\bibnamefont {Morandotti}},
  \bibinfo {author} {\bibfnamefont {G.}~\bibnamefont {Compagno}}, \emph
  {et~al.},\ }\bibfield  {title} {\bibinfo {title} {Direct measurement of
  particle statistical phase},\ }\href@noop {} {\bibfield  {journal} {\bibinfo
  {journal} {ArXiv preprint arXiv:2202.00575 [quant-ph]}\ } (\bibinfo {year}
  {2022})}\BibitemShut {NoStop}%
\bibitem [{\citenamefont {Lo~Franco}(2021)}]{lo2021directly}%
  \BibitemOpen
  \bibfield  {author} {\bibinfo {author} {\bibfnamefont {R.}~\bibnamefont
  {Lo~Franco}},\ }\bibfield  {title} {\bibinfo {title} {Directly proving the
  bosonic nature of photons},\ }\href@noop {} {\bibfield  {journal} {\bibinfo
  {journal} {Nat. Photon.}\ }\textbf {\bibinfo {volume} {15}},\ \bibinfo
  {pages} {638} (\bibinfo {year} {2021})}\BibitemShut {NoStop}%
\bibitem [{\citenamefont {Castellini}\ \emph {et~al.}(2019)\citenamefont
  {Castellini}, \citenamefont {Lo~Franco}, \citenamefont {Lami}, \citenamefont
  {Winter}, \citenamefont {Adesso},\ and\ \citenamefont
  {Compagno}}]{Castellini2019metrology}%
  \BibitemOpen
  \bibfield  {author} {\bibinfo {author} {\bibfnamefont {A.}~\bibnamefont
  {Castellini}}, \bibinfo {author} {\bibfnamefont {R.}~\bibnamefont
  {Lo~Franco}}, \bibinfo {author} {\bibfnamefont {L.}~\bibnamefont {Lami}},
  \bibinfo {author} {\bibfnamefont {A.}~\bibnamefont {Winter}}, \bibinfo
  {author} {\bibfnamefont {G.}~\bibnamefont {Adesso}},\ and\ \bibinfo {author}
  {\bibfnamefont {G.}~\bibnamefont {Compagno}},\ }\bibfield  {title} {\bibinfo
  {title} {Indistinguishability-enabled coherence for quantum metrology},\
  }\href@noop {} {\bibfield  {journal} {\bibinfo  {journal} {Phys. Rev. A}\
  }\textbf {\bibinfo {volume} {100}},\ \bibinfo {pages} {012308} (\bibinfo
  {year} {2019})}\BibitemShut {NoStop}%
\bibitem [{\citenamefont {Sun}\ \emph {et~al.}(2022)\citenamefont {Sun},
  \citenamefont {Liu}, \citenamefont {Wang}, \citenamefont {Hao}, \citenamefont
  {Xu}, \citenamefont {Xu}, \citenamefont {Li}, \citenamefont {Guo},
  \citenamefont {Castellini}, \citenamefont {Lami} \emph
  {et~al.}}]{sun2021experimental}%
  \BibitemOpen
  \bibfield  {author} {\bibinfo {author} {\bibfnamefont {K.}~\bibnamefont
  {Sun}}, \bibinfo {author} {\bibfnamefont {Z.-H.}\ \bibnamefont {Liu}},
  \bibinfo {author} {\bibfnamefont {Y.}~\bibnamefont {Wang}}, \bibinfo {author}
  {\bibfnamefont {Z.-Y.}\ \bibnamefont {Hao}}, \bibinfo {author} {\bibfnamefont
  {X.-Y.}\ \bibnamefont {Xu}}, \bibinfo {author} {\bibfnamefont {J.-S.}\
  \bibnamefont {Xu}}, \bibinfo {author} {\bibfnamefont {C.-F.}\ \bibnamefont
  {Li}}, \bibinfo {author} {\bibfnamefont {G.-C.}\ \bibnamefont {Guo}},
  \bibinfo {author} {\bibfnamefont {A.}~\bibnamefont {Castellini}}, \bibinfo
  {author} {\bibfnamefont {L.}~\bibnamefont {Lami}}, \emph {et~al.},\
  }\bibfield  {title} {\bibinfo {title} {Activation of
  indistinguishability-based quantum coherence for enhanced metrological
  applications with particle statistics imprint},\ }\href@noop {} {\bibfield
  {journal} {\bibinfo  {journal} {PNAS}\ }\textbf {\bibinfo {volume} {119}},\
  \bibinfo {pages} {e2119765119} (\bibinfo {year} {2022})}\BibitemShut
  {NoStop}%
\bibitem [{\citenamefont {Perez-Leija}\ \emph {et~al.}(2018)\citenamefont
  {Perez-Leija}, \citenamefont {Guzm{\'a}n-Silva}, \citenamefont
  {Le{\'o}n-Montiel}, \citenamefont {Gr{\"a}fe}, \citenamefont {Heinrich},
  \citenamefont {Moya-Cessa}, \citenamefont {Busch},\ and\ \citenamefont
  {Szameit}}]{perez2018endurance}%
  \BibitemOpen
  \bibfield  {author} {\bibinfo {author} {\bibfnamefont {A.}~\bibnamefont
  {Perez-Leija}}, \bibinfo {author} {\bibfnamefont {D.}~\bibnamefont
  {Guzm{\'a}n-Silva}}, \bibinfo {author} {\bibfnamefont {R.~d.~J.}\
  \bibnamefont {Le{\'o}n-Montiel}}, \bibinfo {author} {\bibfnamefont
  {M.}~\bibnamefont {Gr{\"a}fe}}, \bibinfo {author} {\bibfnamefont
  {M.}~\bibnamefont {Heinrich}}, \bibinfo {author} {\bibfnamefont
  {H.}~\bibnamefont {Moya-Cessa}}, \bibinfo {author} {\bibfnamefont
  {K.}~\bibnamefont {Busch}},\ and\ \bibinfo {author} {\bibfnamefont
  {A.}~\bibnamefont {Szameit}},\ }\bibfield  {title} {\bibinfo {title}
  {Endurance of quantum coherence due to particle indistinguishability in noisy
  quantum networks},\ }\href@noop {} {\bibfield  {journal} {\bibinfo  {journal}
  {npj Quant. Inf.}\ }\textbf {\bibinfo {volume} {4}},\ \bibinfo {pages} {45}
  (\bibinfo {year} {2018})}\BibitemShut {NoStop}%
\end{thebibliography}
%

\end{document}